\documentclass[12pt]{spieman}% 12pt font required by SPIE;

\usepackage{amsmath,amsfonts,amssymb}
\usepackage{graphicx}
\usepackage{setspace}
\usepackage{tocloft}
\usepackage{xcolor}
\usepackage{lineno}
\title{Recovering the Dust Mass Budget with PRIMA}

\author[a]{A. Traina}
\author[a,b]{F. Pozzi}
\author[a]{F. Calura}
\author[a,b]{M. Costa}
\author[c]{L. Bisigello}
\author[a]{C. Gruppioni}
\author[d,e,f]{L. Barchiesi}
\author[a]{I. Delvecchio}
\author[a]{L. Vallini}
\author[a,b]{C. Vignali}
\author[a,f]{V. Casasola}

\affil[a]{INAF -- Osservatorio di Astrofisica e Scienza dello Spazio di
Bologna, via Gobetti 93/3, 40129, Bologna, Italy}
\affil[b]{Dipartimento di Fisica e Astronomia, Università of Bologna, via Gobetti 93/2, 40129, Bologna, Italy}
\affil[c]{Dipartimento di Fisica e Astronomia, Università of Padova, via F. Marzolo, 8, 35131 Padova, Italy}
\affil[d]{Department of Astronomy, University of Cape Town, 7701 Rondebosch, Cape Town, South Africa}
\affil[e]{Inter-University Institute for Data Intensive Astronomy, Department of Astronomy, University of Cape Town, 7701 Rondebosch, Cape Town, South Africa}
\affil[f]{INAF–Istituto di Radioastronomia, Via Piero Gobetti 101, I-40129 Bologna, Italy}

\cftpagenumbersoff{figure}
\cftpagenumbersoff{table} 
\begin{document} 
\maketitle

\begin{abstract}
Achieving a complete picture of galaxy evolution is a primary goal of extragalactic astrophysics. To accomplish this ambitious task, a wealth of multi-wavelength surveys have been devoted to assess the cosmic evolution of the cold gas and of the stellar mass across cosmic time. In this cosmic census, one elusive component is represented by interstellar dust. In this work, we exploit the IR mission PRIMA (covering wavelengths from 24 $\mu$m to 235 $\mu$m) to perform a deep survey (1000h on 1 deg$^2$) aimed at estimating the still poorly known dust mass function (DMF) at $z \sim 0.5 - 5$.
%, i.e. over a major fraction of cosmic time. 
We consider the spectro-photometric realization of the SPRITZ simulation and we compute the dust masses using single temperature Modified Grey Body functions. We show how PRIMA alone, thanks to its unprecedented sensitivities, will constrain the DMF at $z < 1.5$, in terms of mass and faint-end slope.
%, sampling the dust masses down to 10$^7$ M$_{\odot}$, at least an order of magnitude fainter than previous determinations. 
At $z > 1.5$, we stress the key synergy with current or future sub-millimeter facilities, such as the JCMT/SCUBA-2, AtLAST, LMT and ALMA telescopes, that will allow us to probe the R$-$J regime of PRIMA selected galaxies.
%up to $z \sim 5 \, - \, 6$.
Finally, PRIMA, thanks to its large photometric coverage,
%number of photmetric points in the 24-230 $\mu$m range,
will be able for the first time to constrain strictly the warm dust properties of a two component dust model.

\end{abstract}

% Include a list of up to six keywords after the abstract

% Include email contact information for corresponding author
\noindent \footnotesize\textbf{*}First author mail:  \linkable{alberto.traina@inaf.it}

\begin{spacing}{2}   % use double spacing for rest of manuscript

\section{Introduction}
\label{sect:intro}  % \label{} allows reference to this section

One major goal of extragalactic astronomy is to trace in the most complete possible way the evolution of baryonic matter across cosmic time. One fundamental component of galaxies is interstellar dust, representing the solid component of the interstellar medium and affecting the spectral properties of galaxies over a wide range of wavelengths, ranging from the far-infrared to the ultraviolet domain (e.g.\cite{draine2009dust}), playng a key role in shaping their evolution (\cite{mckee2007sf,cazaux2002ism,hopkins2012stellarfeedback}). How did the dust mass budget evolve through cosmic time? Addressing this question is of the utmost importance to constrain one significant component of the cold mass fraction in galaxies (e.g. \cite{li2004ASSL..319..535L,santini2014A&A...562A..30S,casasola2017A&A...605A..18C}), to access obscured star formation (e.g., \cite{blain1996tdust,gruppioni2017PASA...34...55G,algera2023MNRAS.518.6142A}) and the fraction of heavy elements removed from the gas phase and incorporated into solid grains (e.g., \cite{calura2008A&A...479..669C,jenkins2009ApJ...700.1299J,vladilo2011A&A...530A..33V,Konstantopoulou2022A&A...666A..12K}). A fundamental quantity suited for this purpose is the dust mass function (DMF). A thorough estimate of the evolution of the DMF will allow us to reconstruct how the production and destruction of interstellar dust have changed over time in galaxies of various masses. Thus far, the local DMF has been recovered by various authors (e.g. \cite{vlahakis2005dmd,clements2010dust,beeston2018dmd}), while the evolution of the DMF, based on FIR/sum-mm selection, has been subject of a handful of studies. \cite{dunne2011dmd} considered a sample of 250 $\mu$m-selected sources from the {\it Herschel}-ATLAS Science Definition Phase to look at the evolution of galaxies' space density as a function of their dust mass from $z=0$ up to redshift $z{\sim}$ 0.5. 
Their results supported a sharp increase in the bright end of the DMF across this redshift interval and it has been recently confirmed by \cite{Beeston24} using a sample with an order of magnitude more galaxies than used in previous analyses. 

%consistent with a previously determined DMF estimate at $z \sim 2.5$ based on a limited sample of SCUBA sub-millimetre galaxies (Dunne et al. 2003 \cite{dunne2003dmf}). 
\par In \cite{pozzi2020dmf}, the first study of the DMF evolution from z ${\sim}$ 0.2 up to z ${\sim}$  2.5 has been presented, based on a far-IR (FIR, 160 ${\mu}$m) {\it Herschel}-selected catalogue in the COSMOS field. The sample of \cite{pozzi2020dmf} consisted of ${\sim}$ 5500 sources with flux density $>$ 16 mJy and estimated spectroscopic or photometric redshift. For each of these systems, the dust mass $M_{\rm D}$ was derived from the observed flux by assuming a modified black-body relation, valid for a single-temperature ($T_{\rm D}$) dust component and in the standard optically thin regime (\cite{bianchi2013dust,hunt2019dust}). This estimate requires the assumption of a temperature value, which comes from an empirical relation between $T_{\rm D}$, SFR and redshift out to $z{\sim}$ 2 (\cite{magnelli2014dust}).
However, the observations performed with {\it Herschel} suffer from severe limitations due mostly to its poor sensitivity and resolution, which caused significant confusion noise and allowed one to probe only the bright-end ($M_{\rm DUST} > 10^{9}$ M$_{\odot}$) of the DMF at $1<z<2$. At $z>3$, there is only one ALMA-based exploration, from \cite{traina2024dmf}. In this work, the authors used ${\sim}$200 serendipitously detected galaxies from the ALMA A$^3$COSMOS database (\cite{liu2019a31,adscheid2024a3cosmos}) to sample the DMF at $z = 0.5-6$. In the redshift interval in common with \cite{pozzi2020dmf} (0.5 $<z<$ 2.5), the authors find consistent results but, again, in each redshift bin they were able to sample only the larger dust masses ($M_{\rm DUST} = 10^{9}-10^{10}$ M$_{\odot}$). Thus, a significant component of the DMF (i.e., the faint-end, $M_{\rm DUST} \lesssim 10^{8}$ M$_{\odot}$ ) is currently missing in high-redshift estimates.     
\par Current cosmological simulations allow one to track the evolution of basic galaxy properties, such as star formation, metal content and dust mass (\cite{parente22,celine24,popping2017dmd_sim,aoyama2018dmd_sim,li2019dmd_sim,vijayan2019dmd_sim}). Some current models underpredict the DMF already at $z \sim 1$, fail to account for the characteristic mass or lack the brightest objects (\cite{hou2019MNRAS.485.1727H}). The present poor theoretical understanding of the evolution of the dust mass budget outlines the need for pursuing more surveys at high redshift, to better constrain the shape of the DMF and achieve new estimates at the so-called “Cosmic Noon”, i.e. at $1<~z<3$, where the measured peak of cosmic star formation lies. In this paper, we will explore the capabilities of The PRobe far-Infrared Mission for Astrophysics, PRIMA\footnote{https://prima.ipac.caltech.edu/} (PI: GJ. Glenn) to achieve a significant progress in the understanding of the dust mass budget evolution as a function of cosmic time. PRIMA is currently in the mission concept phase, with potential approval in 2026.  PRIMA is the only instrument that will enable a blind, therefore unbiased IR survey, and thanks to its sensitivity  will overcome the limitations suffered by {\it Herschel}. In this paper we will show indeed the predictions on the dust mass range and DMF as probed by PRIMA, using the $60\, \mu$m PRIMAger detected sources to the overcome confusion limit at longer wavelengths.
With its unprecedented features, PRIMA will:

\begin{itemize}
    \item allow us to improve considerably the characterisation of the DMF and the basic parameters that define its shape, enabling a better estimate of the faint end and the sampling of the characteristic dust mass across a wide redshift range;
    \item extend the redshift range where the dust emission in individual galaxies can be probed, allowing us to derive the DMF, with PRIMA alone data, up to $z=1.5$ and, in synergy with ALMA, up to $z=5$
    \item allow us to significantly constrain dust production in galaxy formation models, improving our poor theoretical knowledge of this process and filling the gap in our understanding of dust-obscured galaxies.
\end{itemize}

\section{PRIMA Dust Mass Function}

\subsection{PRIMA}
PRIMA is a proposed far-infrared observatory, equipped with a 1.8m diameter mirror and able to perform imaging and spectroscopic studies in the 24$-$235 $\mu$m range.  PRIMA will be equipped with two instruments: the spectrograph FIRESS and  the multi-band spectro-photometric imager PRIMAger. We refer to the PRIMA science book\cite{prima_sciencebook2023} for an exhaustive description of the two instruments, while here we remind the main characteristics of PRIMAger, the best instrument among those on board of PRIMA to perform cosmological survey given its high mapping speed, $\sim 2$dex better than {\it Herschel}. PRIMAger consists of two hyperspectral band with variable filters (R=10, PHI1 and PHI2) in the 24$-$80 $\mu$m range, and 4 broad bands filters (PPI1, PPI2, PPI3, PPI4) in the 80$-$230 $\mu$m range. In the following, we will refer only to the total intensity capabilities of PRIMA and we will consider the two hyperspectral band PHI1 and PHI2 divided in 12 filters.

\subsection{Simulated PRIMA galaxies and dust mass estimates }
\label{simulation_sec}

\begin{figure}
\begin{center}
\includegraphics[height=10.5cm]{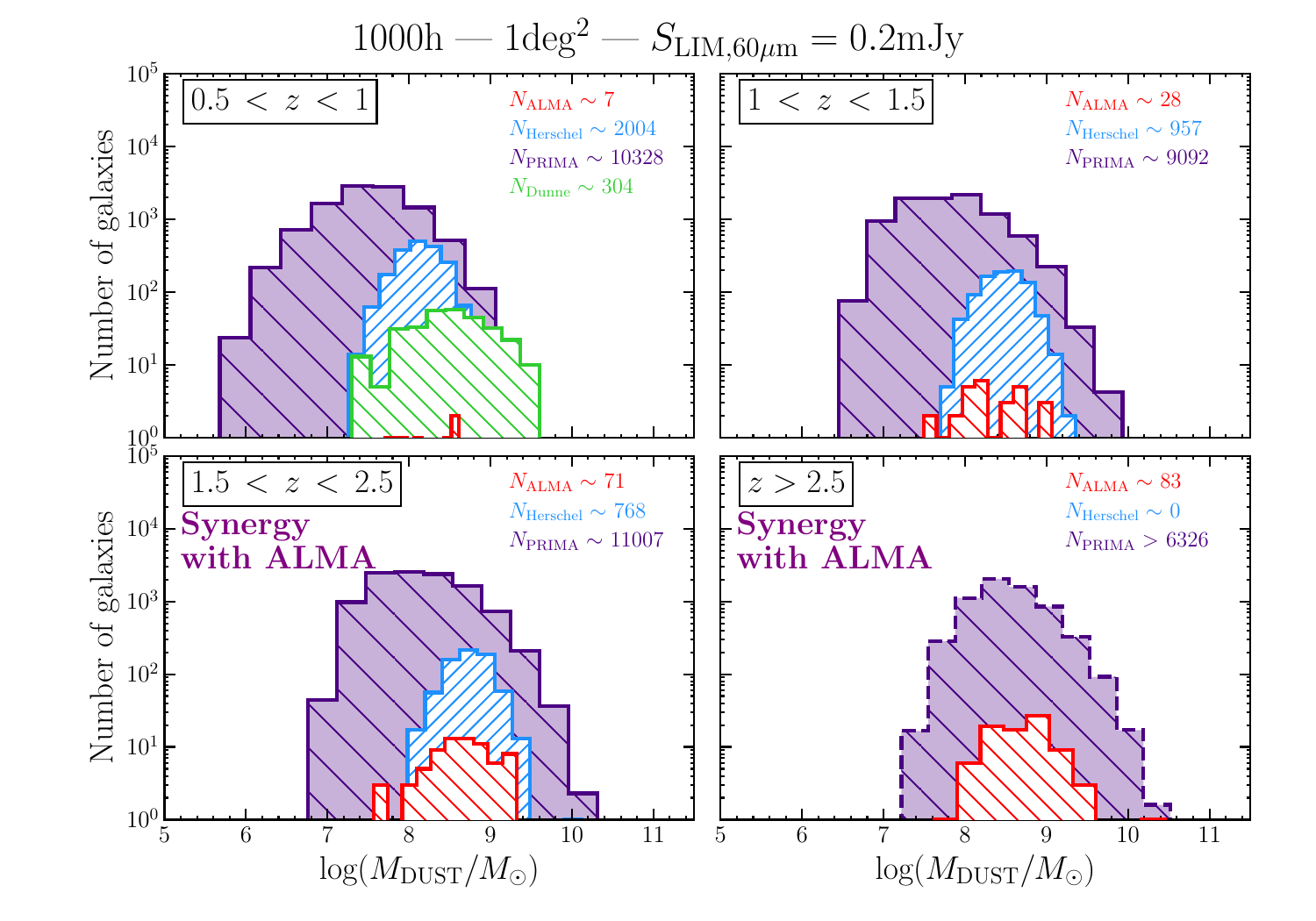} 
\caption{Dust mass distribution of the galaxies used to derive the DMFs with {\it Herschel} (\cite{pozzi2020dmf} blue), ALMA (\cite{traina2024dmf} red) and PRIMA (purple). The distribution of $M_{\rm DUST}$ probed by PRIMA is obtained by leveraging the SPRITZ simulation (\cite{bisigello2021spritz}). As a comparison, the data from \cite{dunne2011dmd} are reported in the top left panel in green.}
\label{fig:dust_distribution}
\end{center}
\end{figure}

To study the improvement achievable on the characterization of DMF, we consider a deep “reference survey” covering 1 deg$^2$, with a total integration time of 1000hr (overhead included). By this strategy, we achieve 5$\sigma$ sensitivities in the range $0.09-0.23$ mJy within the 24$-$230 $\mu$m PRIMAger wavelength interval (see the PRIMA fact sheets \hyperlink{https://prima.ipac.caltech.edu/page/fact-sheet}{https://prima.ipac.caltech.edu/page/fact-sheet}).

To calculate the expected number of detectable galaxies with PRIMAger, we use the spectrophotometric realizations of IR-selected targets at all-z, SPRITZ  (\cite{bisigello2021spritz,bisigello2024spritz}). The SPRITZ simulations are based on a phenomenological model whose starting point is the observed IR luminosity function obtained with  {\it Herschel} data (\cite{gruppioni2013lf}). The SPRITZ model includes different population of IR active galaxies, such as star-forming
galaxies, AGN, and composite systems, along with elliptical
galaxies and  dwarf irregular galaxies. The SPRITZ simulation produces a catalog of simulated SEDs (each associated to a specific template), with simulated PRIMA photometry. In our analysis, we used these SEDs to estimate dust masses and temperatures.

Using the SPRITZ realizations, considering the strategy described above, we expect to detect $\sim 44000$ galaxies with fluxes  above the limiting flux of 0.2 mJy in the 9$^{\text{th}}$ filter (PHI2 at $\sim$ 60 $\mu$m). We consider the 60 $\mu$m filter as the “detection” filter since it provides the highest observed IR flux, given an IR spectral energy distribuion (SED), without reaching the confusion limit (see \cite{bethermin24} for details on the confusion noise) and the benefit of the rapid rise of the IR SED. We plan to use the 60 $\mu$m \footnote{orlower wavelengths} positions as priors to deblend the longer wavelengths (see \cite{donnellan2024confusion}). Considering the typical galaxy SEDs of the SPRITZ realizations, we infer a typical 160/60 flux ratio of approximately 10 in a large redshift interval (0.5$ < z <$ 5, decreasing to ${\sim}$ 3 towards the local Universe). This implies a 160 $\mu$m detection limit of 2 mJy, nearly 1 dex deeper than {\it Herschel}.

We derive $M_{\rm D}$ of the simulated galaxies using a single temperature Modified Black Body (MBB) in the optically thin approximation (see \cite{magnelli2020dmd}):

\begin{equation} \label{eq:mdust}
 \centering
    M_{\rm D} = \frac{5.03 \times 10^{-31} \cdot S_{\nu_{\rm obs}} \cdot D_{\rm L}^2}{(1+z)^4 \cdot B_{\nu_{\rm obs}}(T_{\rm obs}) \cdot \kappa_{\nu_o}} \cdot \left( \frac{\nu_o}{\nu_{\rm rest}}  \right)^{\beta} ;
\end{equation}

where $ B_{\nu_{\rm obs}}(T_{\rm obs})$ is the black-body Planck function computed at the observed frame, mass-weighted, temperature ($T_{\rm obs} = T_{\rm D} / (1+z)$);  $S_{\nu_{\rm obs}}$ is the flux density at the observed frequency $\nu_{\rm obs}$, with $\nu_{\rm rest} = (1+z) \nu_{\rm obs}$; $\kappa_{\nu_o}$ is the photon cross-section to mass ratio of dust at the rest-frame $\nu_0$, and $\beta$ is the dust emissivity spectral index. Here we used $\beta = 1.8$, as suggested value by \cite{scoville2014dust} based on the findings of the \cite{planck_collab2011} $\kappa = 0.0469$ m$^2$kg$^{-1}$, derived for a wavelength of $850 \mu$m (\cite{draine2014dustem}). The dust temperature has been computed for each galaxy following the empirical relation by \cite{magnelli2014dust}:
\begin{equation}
    T_{\rm D} = 98 \times (1+z)^{-0.065} + 6.9 \times {\rm log}(SFR/M_{\star})
\end{equation}
The applicability of the optically thin approximation is valid in the Rayleigh–Jeans (R$-$J) regime. This point, along with the possibility of directly deriving the dust temperature $T_{\rm D}$ from the shape of the SED (see also Sect. \ref{temp_sec}), leads us to adopt two different approaches depending on the galaxy redshift, $z$: for galaxies at $z < 1.5$, we solely rely on the PRIMAger simulated fluxes, and we use the flux at the longest wavelengths band (230$\mu$m); for galaxies at $z > 1.5$, instead, we extrapolate the expected fluxes at $\lambda > 230 {\mu}$m from the simulated SED available in the SPRITZ realizations. This highlights how, at
$z \sim 1.5$,  the synergy between PRIMA and a sub-mm/mm telescope such as, for example, JCMT/SCUBA-2, AtLAST, LMT or ALMA will be fundamental. Indeed, these facilities will allow us to probe the R-J part of the SED of galaxies selected by PRIMA at shorter wavelength. Given the evergrowing nature of the ALMA archive, we expect the number of PRIMA selected galaxies, with an ALMA counterpart, to reach large statistics in the next $\sim 10$ years (i.e., corresponding to the possible launching date of PRIMA). Other future facilities (e.g., the AtLAST single-dish telescope), will have a fundamental role in probing the R-J part of the SED for a large number of galaxies in relatively short times (e.g., the estimated FoV of AtLAST is expected to be $\sim 1$ deg), reaching limiting fluxes small enough to detect PRIMA-selected galaxies, with an observing time comparable with PRIMA.
In Figure \ref{fig:dust_distribution}, we show the dust mass distribution in 4 redshift bins that PRIMA will be able to recover with an integration time of 1000h on 1 deg$^2$. Together with the PRIMA expectations, the distribution of the dust masses used for the DMF derived with {\it Herschel} (\cite{pozzi2020dmf}) and ALMA (\cite{traina2024dmf}) are shown. In all redshift intervals, the number of sources that will be used for the DMF will be at least 1 dex higher than in previous {\it Herschel} surveys at 1.5 $< z < $ 5. Moreover, it is evident how the sensitivities of PRIMA will allow us to extend the mass range of the DMF sampled, especially in the faint-end, down to $10^{6}$ M$_{\odot}$.

\subsection{Comparison with previous DMFs estimates}
In the past years, several works have been carried out with the goal of deriving the dust properties in the local, as well as in the high-redshift Universe, several of which were able to estimate the DMF at $0 < z <3$ (\cite{dunne2003dmf,magnelli2019dmd,pozzi2020dmf,traina2024dmf}).
However, except for the lowest redshift bins, even current works are not able to trace the low-mass end of the DMF. This lack of data in the faint-end (i.e., $M_{\rm D} < 10^{8}$ M$_{\odot}$) at $z \geq 0.5$ translates into an extrapolation of the Schechter function and into large uncertainties on the population of galaxies with low dust content. Thanks to its photometric coverage and to its unprecedented sensitivity, PRIMA is perfectly suited to probe the lowest masses, allowing us to tighten the currently large uncertainties. 
\par
\begin{figure*}
\begin{center}
\includegraphics[height=10.5cm]{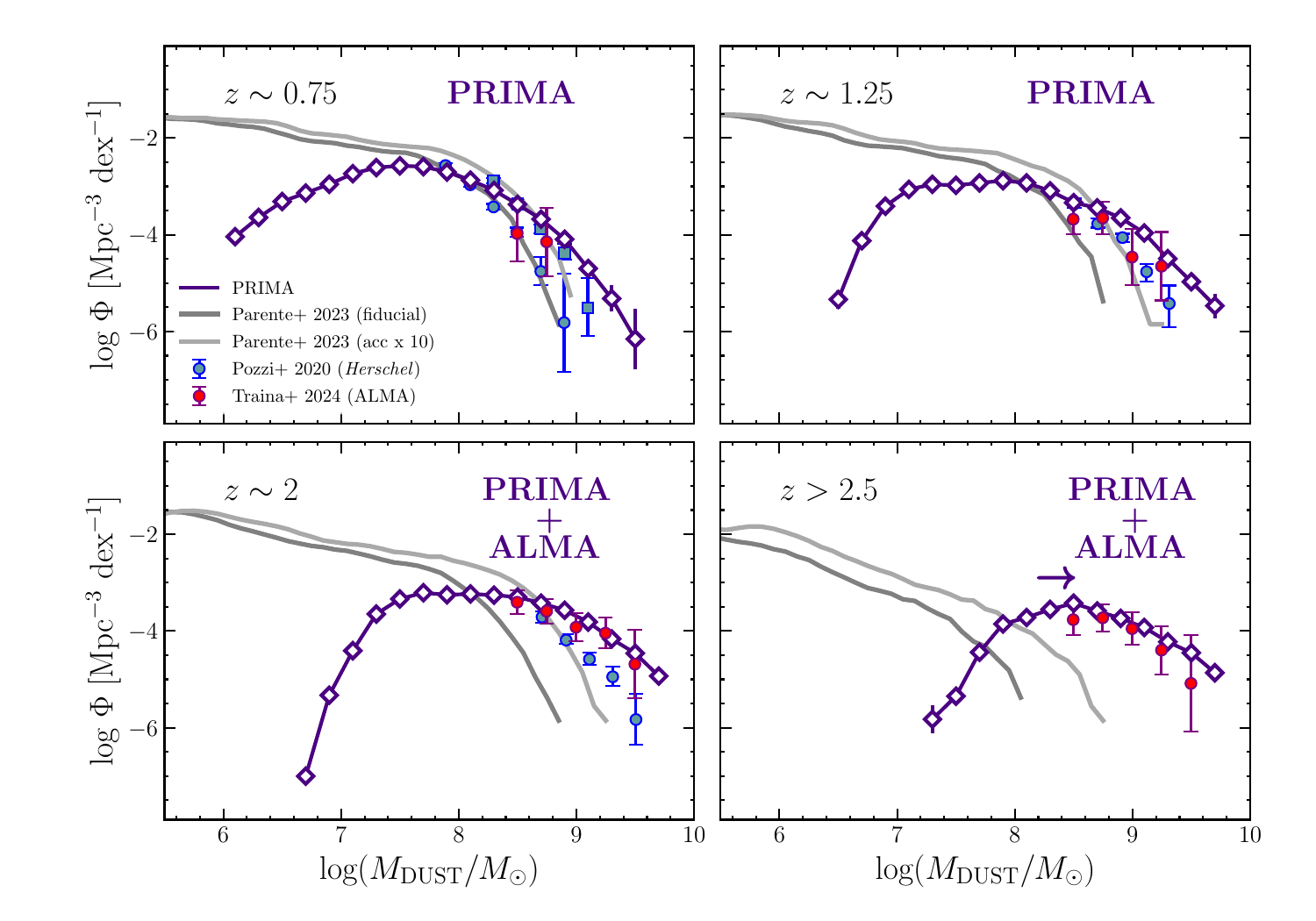}
\end{center}
\caption{Dust mass functions at $z \sim 0.75$, $z \sim 1.25$, $z \sim 2$ and $z > 2.5$ from models and literature data, compared with the dust mass range probed by PRIMA. The dark and light grey lines are two different realization of the SAM by \cite{parente2023dmdm_sim}. The first one corresponds to their fiducial model, while the second one is derived by reducing the accretion timescale by a factor 10, to mimic a more efficient accretion in molecular clouds. Data from \cite{pozzi2020dmf} and \cite{traina2024dmf} are shown as blue and red points, respectively. Blue squares and circles refer to different redshift bins. The PRIMA predicted DMF is plotted as white diamonds and purple lines.} 
\label{fig:PRIMA_DMF}
\end{figure*}

 In the previous section, we described the method to derive the dust masses, from the observed PRIMA flux densities or by extrapolating each object SED, depending on source redshift. In this section, we discuss the comparison between the predicted PRIMA DMF and those derived in other works, as well as the reducing the uncertainties. Figure \ref{fig:PRIMA_DMF} shows the comparison between {\it Herschel} and ALMA DMFs, with the DMF predicted by PRIMA, in four redshift bins ($0.5 < z < 3$). Even the most recent works have been able to constrain the high-mass end only of the DMF, while its knee and faint-end remain unprobed. On the other hand, PRIMA will estimate the faint-end down to $M_{\rm D} \sim 10^6 \, {\rm M_{\odot}}$ at $z \sim 0.75$ and $M_{\rm D} \sim 10^7 \, {\rm M_{\odot}}$ at $0.75 < z < 2$. At very low redshifts (in particular in the first redshift bin), where galaxies with very low dust masses are expected to be detected, we cannot exclude possible effects on the SED shape produced by metallicity.
However, the origin of dust in low-metallicity, low-mass galaxies is still largely unknown and has been investigated in only a limited number of studies, even within the local Universe \cite{fisher2014,lianou2019,calura2021}, making PRIMA even more important in the study of faint sources.
At higher redshifts, it is not possible to sample the R-J part of the SED by relying on PRIMA data alone and ALMA (or other single-dish sub-mm telescopes) data (either proprietary or from the archive) are needed (or extrapolation of the SED at longer wavelengths). The study and characterization of the faint-end of the DMF it is important not only to extend our knowledge to the faintest galaxies, but also to improve the characterization of the physical processes regulating dust formation and evolution in models. In Figure \ref{fig:PRIMA_DMF} two sets of semi analytical models (SAMs) by \cite{parente2023dmdm_sim}, with different dust production efficiency, are shown. Even though the characterization of the bright-end can help in discriminating between models, however, without a proper sampling of the faint-end, it is not possible to fully reconstruct dust properties form models. 
\par
\begin{figure*}
\begin{center}
\includegraphics[height=10.5cm]{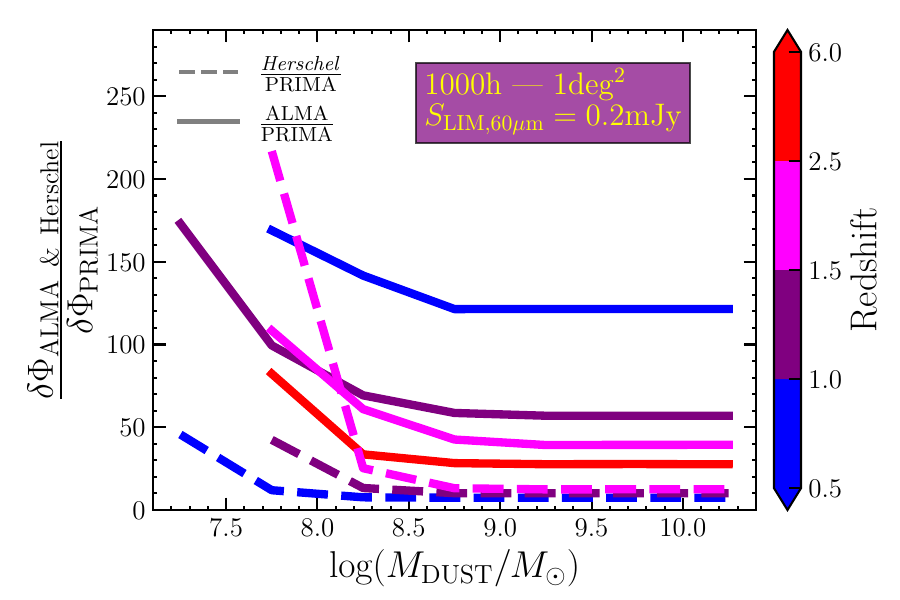}
\end{center}
\caption{Ratios between the uncertainties on the DMF as derived by {\it Herschel} (dashed line) or ALMA (solid line) and the prediction with PRIMA. Different colors represent different redshift ranges.
} 
\label{fig:DMF_uncertainties}
\end{figure*}

 The major strengths of PRIMA for the DMF estimate are the wider range covered in wavelength and the larger statistics. From the numbers reported in Figure \ref{fig:dust_distribution}, we derived the expected improvements in terms of uncertainties in the DMF, compared to the estimates from {\it Herschel} and ALMA (see Figure \ref{fig:DMF_uncertainties}). Assuming that the Poissonian uncertainties on the DMF scale as $N^{-2}$, where $N$ is the number of objects, PRIMA will improve the error estimates by a factor of $50-150$ with respect to current ALMA surveys and up to 50 with respect to current {\it Herschel} surveys, up to $z \sim 1.5$. At high redshifts, thanks to the synergy with ALMA, the errors will be improved by a factor of $50-100$. Figure \ref{fig:DMF_uncertainties} shows how these uncertainties will be reduced, as a function of the dust mass.

\subsection{Synergy with current and future (sub-)mm facilities}
\begin{figure*}
\begin{center}
\includegraphics[height=10.5cm]{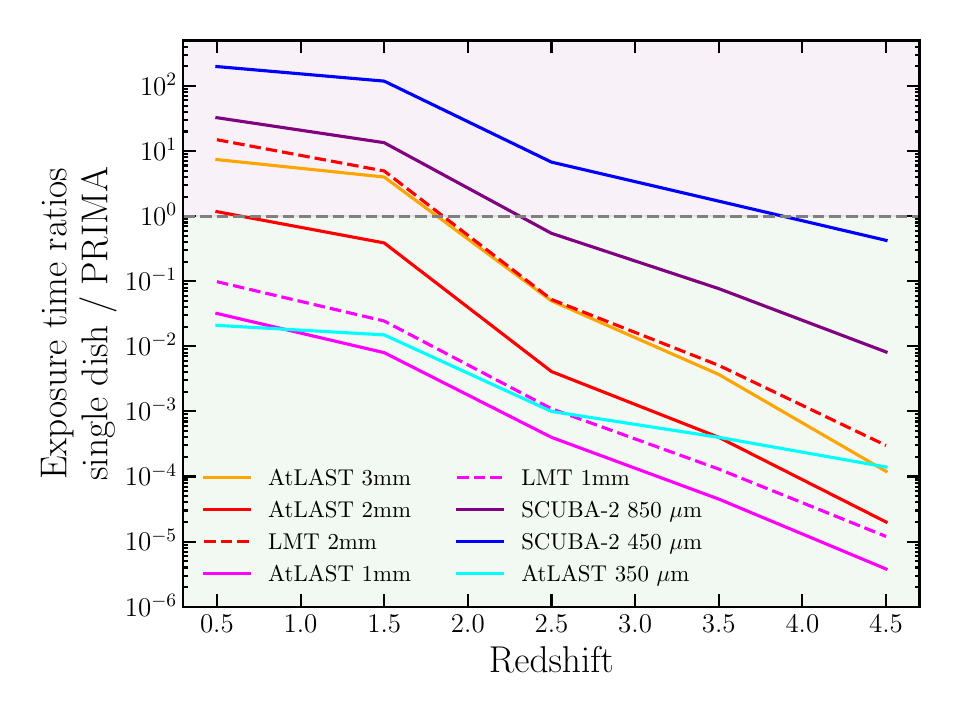}
\end{center}
\caption{Ratios between the exposure times of single-dish (sub-)mm facilities and PRIMA, on 1 deg$^2$. Lines with different colors represents different wavelengths, while different linestyles are use for different telescopes when plotting curves at the same wavelength. The pink shaded area represents the region in which the exposure times are lower with PRIMA, while in the light-green area single-dish telescopes reach the required sensitivity with lower exposure times.}
\label{fig:exptimes}
\end{figure*}

As we already mentioned, PRIMA will be able to probe the R-J part of the SED up to $z \sim 1.5\,-\,2$. At higher redshift, the lack of coverage of the R-J may lead to inaccurate estimate of the dust masses. For this reason, the synergy of PRIMA with current of future sub-mm and mm facilities (e.g., JCMT/SCUBA-2, LMT, AtLAST) will be of key importance. In this Section, we show how these single-dish telescopes will provide observations on a sky area compatible with the PRIMA 1deg$^2$ survey, with lower exposure times (at $z > 2$).
From the SPRITZ SED, we derived the flux ratios between the 60 $\mu$m and other wavelength at $350\, {\rm \mu m} \,- \lambda\,-\,3\,{\rm mm}$, to obtain the sensitivity needed to detect a 60 $\mu$m PRIMA detected source at longer wavelengths. From these ratios, using the online ETC for JCMT/SCUBA-2, LMT and AtLAST, we computed the exposure times required to observe 1 deg$^2$ reaching the required sensitivity. The results are shown in Figure \ref{fig:exptimes}. At $z >2$ most of these facilities will be able to measure a flux on the R-J and, therefore, will be crucial in the derivation of the dust masses at those redshifts.

\section{Constraining the dust temperature}
\label{temp_sec}

In Section \ref{simulation_sec}, we simulate the dust masses by assuming that the FIR emission can be accurately represented by an isothermal Modified Black Body (MBB) function, within the optically thin regime. This simplification provides a good estimate of the bulk of the dust mass, as shown by the comparison between the dust masses derived from the MBB function and those obtained using more complex dust emission models (see \cite{bianchi2013dust,hunt2019dust}).
\par
\begin{figure}
\begin{center}
\includegraphics[width=1.1\textwidth]{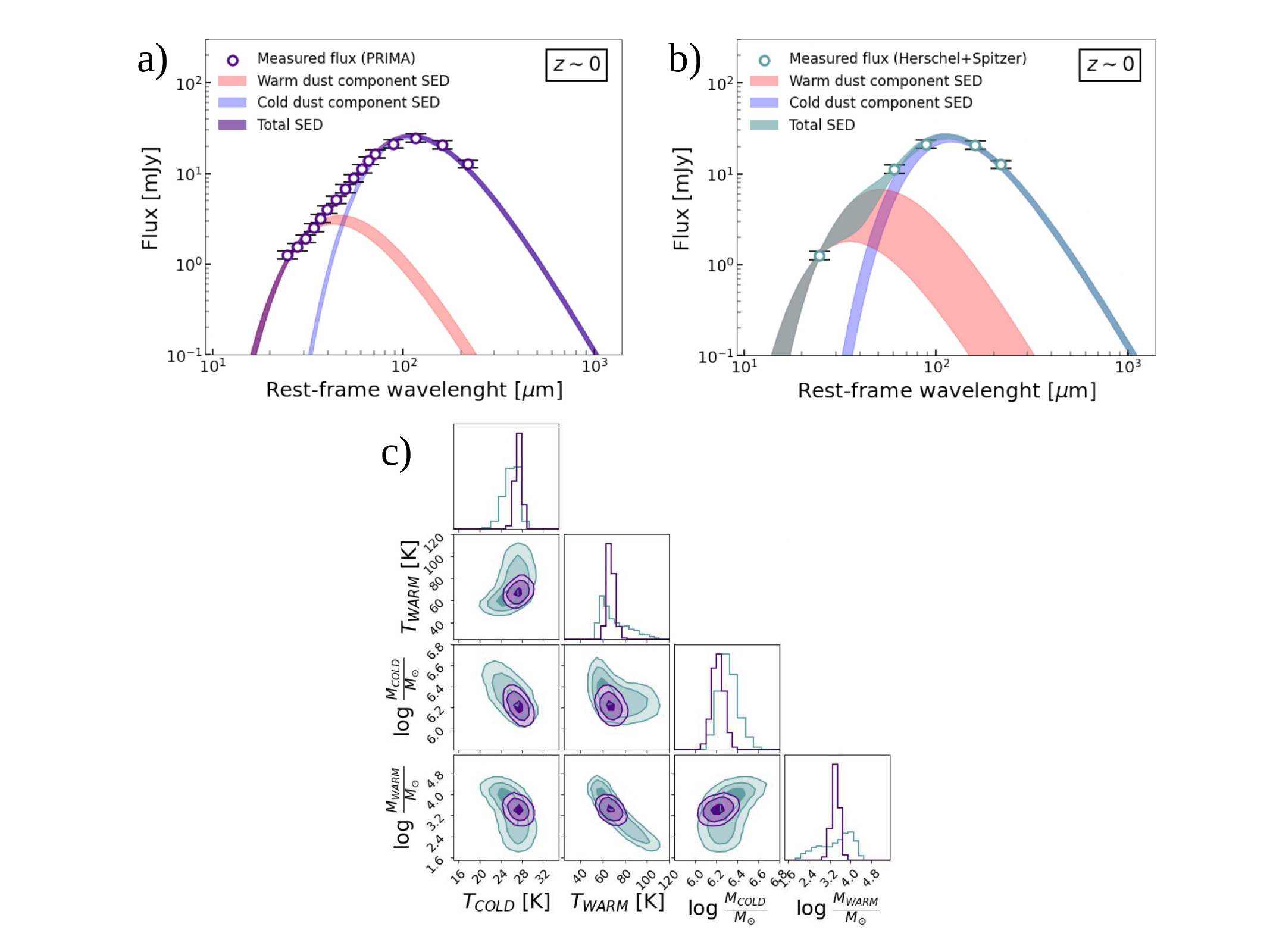}
\end{center}
\caption{Simulated photometry of two MS$-$galaxies at $z=0$ (panels a, b) from the SPRITZ realization (\cite{bisigello2021spritz}). In panels a), b) are displayed the pre-PRIMA and PRIMA photometry, respectively. The best-fitting two$-$dust component model is shown in each panel with the associated 1$\sigma$ uncertainties (red: warm component; blue: cold component). In panels c) the posterior probabilities of the dust parameters using the pre-PRIMA and PRIMA photometry are shown (panel c; blue and violet contours, respectively).}
\label{fig:temp_prima_herschel}
\end{figure}

\begin{figure}
\begin{center}
\includegraphics[width=\textwidth]{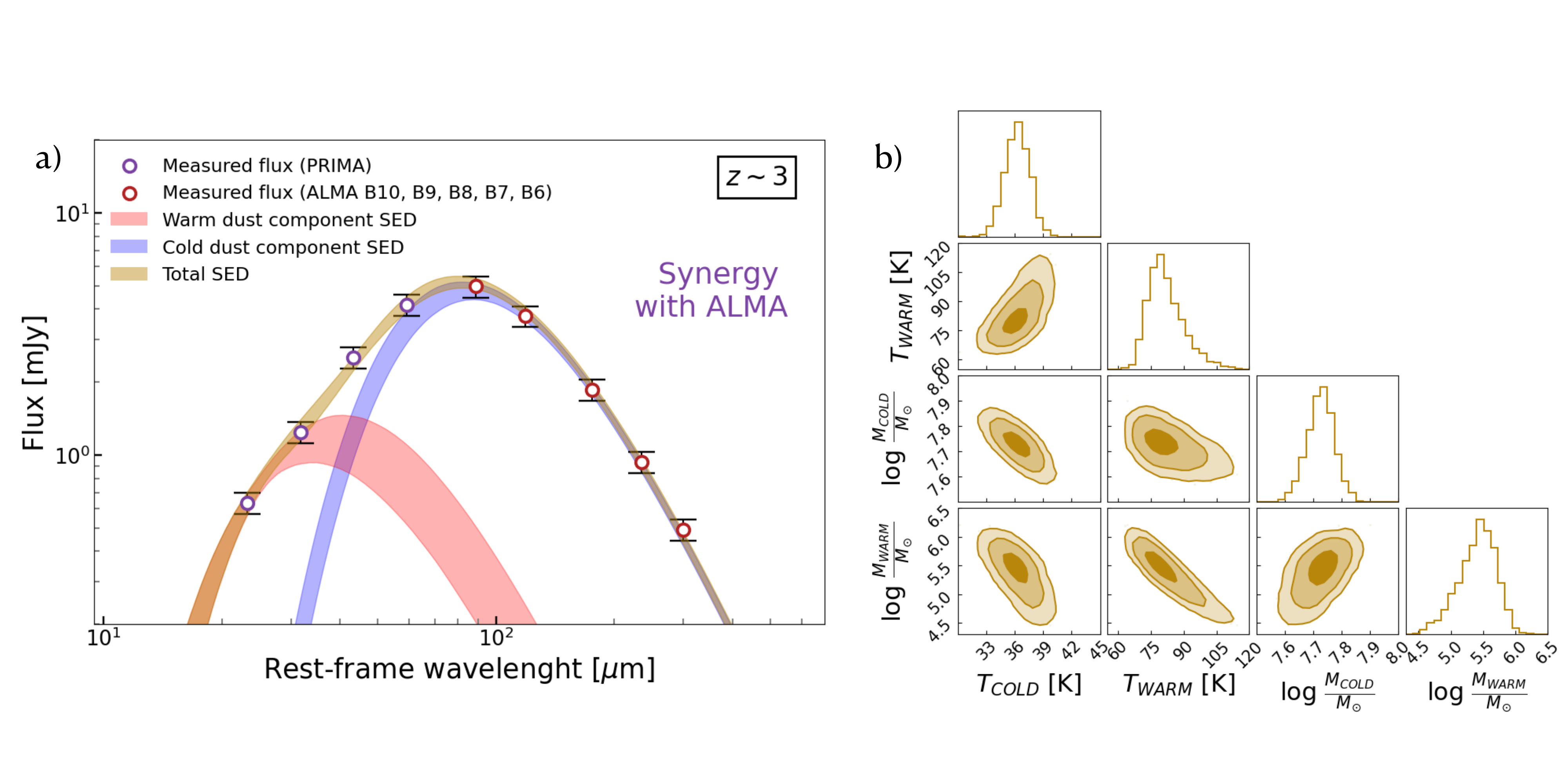}
\end{center}
\vspace{-1.cm}
\caption{Simulated PRIMA+ALMA photometry of a galaxy at $z=3$ (panel a) from the SPRITZ realization (\cite{bisigello2021spritz}). We display the PRIMA points down to their detection limit (i.e., the other PRIMA bands are below the detection limit).The best-fitting two$-$dust component model is shown in panel a) with the associated 1$\sigma$ uncertainties (red: warm component; blue: cold component). In panels b) the posterior probabilities of the dust parameters using the PRIMA+ALMA photometry are displayed.}
\label{fig:temp_prima_alma}
\end{figure}

 In reality, dust grains exhibit a broad range of temperatures, which are determined by both their size distribution and the intensity of the radiation field that heats them (\cite{draine2003dust}). In general, a two-temperature model captures well the diversity in dust temperatures (see \cite{galametz2012tdust,orellana2017dust_prop}). The two dust components are a warm one (with $20 < T < 60$ K), typically associated with photodissociation regions (PDRs), and a cold one (with $T < 30$ K) that is primarily linked to the diffuse interstellar medium (ISM) (see \cite{draine2007dust}). Figures \ref{fig:temp_prima_herschel} and \ref{fig:temp_prima_alma} show the ability of PRIMAger to constrain the dust masses and temperatures assuming a two-temperatures model. In this case the two components are modeled by two MBB functions. In order to compute their dust masses from the relation $\tau_\nu = k_\nu\times \Sigma_{\text{dust}}$ (with $\tau_\nu$ being the optical depth), we assumed a power law behavior of the opacity $k_\nu$ (\cite{carniani2019MNRAS.489.3939C}) and a physical size of 10 kpc.
 The best fit parameters are retrieved  via Markov Chain Monte Carlo (MCMC) fitting implemented with \texttt{emcee} (\cite{foremanmackey2013emcee}). In the top panels of Figure \ref{fig:temp_prima_herschel}, we show the simulated photometry of a Main Sequence galaxy at $z =0$ from the SPRITZ  realization. Panel a) displays the PRIMAger photometric points, and in panel b) we show the simulated pre-PRIMA photometric points ({\it Spitzer}/MIPS at 24 $\mu$m, {\it Herschel/PACS} at 65, 96 and 172 $\mu$m, and {\it Herschel/SPIRE} at 250 $\mu$m), which represent the current state-of-the-art. 
 The large number of photometric points provided by PRIMAger in the 24–230$\mu$m range allows for a very precise determination of the two dust components, at least in the local Universe. In particular, considering the warm component, its 1$\sigma$ mass and temperature ranges are at least a factor three smaller than those obtained from Spitzer/24 $\mu$m + {\it Herschel} data (see panel c where the confidence regions are displayed). In Figure \ref{fig:temp_prima_alma} (panel a), we show the simulated photometry of a galaxy at $z = 3$. Here, the key synergy between PRIMAger and ALMA becomes particularly evident, with PRIMAger effectively constraining the warmer dust component, while ALMA is more sensitive to the colder one.

\section{Conclusions}

In this paper, we explore the performance of the PRIMA instrument PRIMAger, in recovering the DMF as a function of the cosmic time. We have considered a “reference survey” covering 1 deg$^2$, with a total integration time of 1000hr. To simulate the galaxies detectable by PRIMAger under this observational strategy, we use the spectro-photometric realization of the SPRITZ code (\cite{bisigello2021spritz, bisigello2024spritz}). The dust masses have been estimated using a single temperature MBB model under the optically thin approximation valid in the R$-$J regime. This led to the adoption of different strategies depending on the redshift $z$ of the sources: for galaxies at $z<1.5$ we are able to properly constrain the dust emission with just PRIMAger photometry, while at $z>1.5$ we rely on the synergies with the ALMA facility. Our main results are as follows:

\begin{itemize}
    \item The sensitivities of the PRIMAger instrument, along with the proposed observational survey, will allow us to detect $\sim 44000$ galaxies and derive the DMF across a wide redshift range of 0$ < z < $5. 
    In particular, up to $z {\sim}$ 2, PRIMAger will detect galaxies that are at least 1 dex less massive than those detectable with current facilities. This will allow for better constraints on the faint-end slope of the DMF, which is currently completely unconstrained at $z > 0.5$.
    
  \item The high number of galaxies that PRIMAger will be able to detect, will reduce the uncertainties on the DMF by at least a factor of 50 compared to {\it Herschel}-based estimates at $z < 2.5$, and by up to a factor of 50 compared to the high$-z$ DMF measurements obtained with ALMA.

  \item We have explored the feasibility of fitting the simulated photometry with two dust components (warm and cold). The large number of photometric points provided by PRIMAger, will allows to recover the parameters (mass and temperature) of both components with significantly higher accuracy, with respect to what was possible with previous facilities (e.g., Spitzer and {\it Herschel}). We stress how, in particular, the warmer component, traced by the PHI1 and PHI2 filters in the 24$-$80 $\mu$m range, will be well constrained for the first time,  by reducing its associated uncertainties by at least a factor of three in comparison to previous determinations.
    
\end{itemize}

\subsection* {Code and data availability}
The simulations (from the SPRITZ code)\cite{bisigello2021spritz} and code developed for the analysis presented in this work are available upon request to the corresponding author. The ancillary data used are from the A$^3$COSMOS database\cite{liu2019a31} and from the {\it Herschel} catalog in the COSMOS field\cite{pozzi2020dmf}.
\subsection* {Disclosures}
The authors declare there are no financial interests, commercial affiliations, or other potential conflicts of interest that have influenced the objectivity of this research or the writing of this paper.

\subsection* {Acknowledgments}
LB acknowledges financial support from the Inter-University Institute for Data Intensive Astronomy (IDIA), a partnership of the University of Cape Town, the University of Pretoria and the University of the Western Cape, and from the South African Department of Science and Innovation’s National Research Foundation under the ISARP RADIOMAP Joint Research Scheme (DSI-NRF Grant Number 150551) and the CPRR HIPPO Project (DSI-NRF Grant Number SRUG22031677). ID acknowledges funding by the European Union – NextGenerationEU, RRF M4C2 1.1, Project 2022JZJBHM: "AGN-sCAN: zooming-in on the AGN-galaxy connection since the cosmic noon" - CUP C53D23001120006
%This unnumbered section is used to identify those who have aided the authors in understanding or accomplishing the work presented and to acknowledge sources of funding. 

%%%%% References %%%%%

\bibliography{report}   % bibliography data in report.bib
\bibliographystyle{spiejour}   % makes bibtex use spiejour.bst

%%%%% Biographies of authors %%%%%
\subsection*{Biographies}
\vspace{2ex}\noindent\textbf{A. Traina} is a postdoc at the INAF-OAS of Bologna. He received his BS and MS degrees in astrophysics from the University of Bologna in 2018 and 2020, respectively, and his PhD degree in astrophysics from the University of Bologna in 2024. His current research interests include galaxies formation and evolution, as well as their interplay with SMBHs.
%\vspace{2ex}\noindent\textbf{First Author} is an assistant professor at the University of Optical Engineering. He received his BS and MS degrees in physics from the University of Optics in 1985 and 1987, respectively, and his PhD degree in optics from the Institute of Technology in 1991.  He is the author of more than 50 journal papers and has written three book chapters. His current research interests include optical interconnects, holography, and optoelectronic systems. He is a member of SPIE.

%\vspace{1ex}
%\noindent Biographies and photographs of the other authors are not available.

\listoffigures
%\listoftables

\end{spacing}
\end{document}